\documentclass{ifacconf}
\pdfoutput=1
\usepackage{xcolor}

\makeatletter
\let\old@ssect\@ssect 
\makeatother

\usepackage{natbib}
\usepackage{hyperref}
\newtheorem{definition}{Definition}
\newtheorem{assumption}{Assumption}
\newtheorem{remark}{Remark}
\newtheorem{lemma}{Lemma}
\usepackage{amsfonts,amssymb}
 
\usepackage{textcomp}
\usepackage{graphicx}
\usepackage{subfig}
\usepackage{courier}
\usepackage{mathtools}
\usepackage{float}
\usepackage{import}
\usepackage{lipsum}
\usepackage{pgfplots}

\DeclarePairedDelimiter{\norm}{\lVert}{\rVert}

\usepackage{tikz}

\DeclareMathOperator*{\argmin}{arg\,min}

\makeatletter
\def\@ssect#1#2#3#4#5#6{%
	\NR@gettitle{#6}
	\old@ssect{#1}{#2}{#3}{#4}{#5}{#6}
}
\makeatother

\begin{document}
	
	\begin{frontmatter}
		
		\title{Data-driven nonlinear predictive control for feedback linearizable systems\thanksref{titlenote}}
		
		\thanks[titlenote]{
			This work has received funding from the European Research Council (ERC) under the European Union’s Horizon 2020 research and innovation programme (grant agreement No 948679). This work was also funded by Deutsche Forschungsgemeinschaft (DFG, German Research Foundation) under Germany’s Excellence Strategy - EXC 2075 - 390740016 and under grant 468094890. We acknowledge the support by the Stuttgart Center for Simulation Science (SimTech).
		}
		
		\author[IRT]{Mohammad Alsalti} 
		\author[IRT]{Victor G. Lopez} 
		\author[IST]{Julian Berberich} 
		\author[IST]{Frank Allgöwer} 
		\author[IRT]{Matthias A. Müller}
		
		\address[IRT]{Leibniz University Hannover, Institute of Automatic Control, 30167 Hannover, Germany. E-mail: \{alsalti, lopez, mueller\}@irt.uni-hannover.de}
		\address[IST]{University of Stuttgart, Institute for Systems Theory and Automatic Control, 70550 Stuttgart, Germany. E-mail: \{julian.berberich, frank.allgower\}@ist.uni-stuttgart.de}
		
		\begin{abstract}
			We present a data-driven nonlinear predictive control approach for the class of discrete-time multi-input multi-output feedback linearizable nonlinear systems. The scheme uses a non-parametric predictive model based only on input {and noisy output data} along with a set of basis functions that approximate the unknown nonlinearities. {Despite the noisy output data as well as the mismatch caused by the use of basis functions}, we show that the proposed multi-step robust data-driven nonlinear predictive control scheme is recursively feasible and renders the closed-loop system practically exponentially stable. We illustrate our results on a model of a fully-actuated double inverted pendulum.
		\end{abstract}
		
		
	\end{frontmatter}
	
	\section{Introduction}
The field of direct data-driven control has recently extensively built on a result from behavioral control theory \citep{Willems05}. This result states that \textit{all} finite-length input-output trajectories of a discrete-time linear time-invariant (DT-LTI) system can be obtained using a linear combination of a single, persistently exciting, input-output trajectory. \textit{Willems' fundamental lemma}, as it is now known, motivated a large number of works on data-based system analysis and (robust) control design for DT-LTI systems (\cite{Markovsky21} and the references therein).\par
Since the fundamental lemma studies finite-length input-output behavior, it is therefore suitable to use as a non-parametric predictive model in receding horizon schemes. For example, in \cite{Yang15, Coulson20}, it was used for data-driven predictive control, for which open loop robustness \cite{Coulson21, Huang22} as well as closed-loop guarantees \cite{Berberich203, Berberich206} have been established. In the above works, the focus was mainly on  DT-LTI systems. In \cite{Berberich204}, a data-driven predictive control scheme for nonlinear systems using successive local linearization was proposed and shown to be stable. By assuming that the dynamics evolve in the dual space of a reproducing kernel Hilbert space (RKHS), \cite{Yingzhao21} propose a data-based predictive controller for classes of nonlinear systems, but no closed-loop guarantees were given.\par
In our previous work \cite{Alsalti2022a}, we presented an extension of the fundamental lemma to the class of DT multi-input multi-output (MIMO) feedback linearizable nonlinear systems. This was done by exploiting linearity in transformed coordinates along with a set of basis functions that depend only on input-output data. Furthermore, we studied the effect of inexact basis function decomposition of the unknown nonlinearities as well as noisy data. Based on these results, we provided methods to approximately solve the simulation and output-matching control problems using input-output data and showed that the difference between the estimated and true outputs is bounded.\par
The contributions of this work are as follows: First, we design a multi-step robust data-driven nonlinear predictive control scheme based on the data-based system parameterization of \cite{Alsalti2022a}. This scheme uses only input-output data and does not require an intermediate model identification step. This is useful for a variety of feedback linearizable systems whose dynamics are unknown (see, e.g., \cite{Murray95differentialflatness}). Second, we prove that the proposed control scheme leads to practical exponential stability of the closed loop despite an inexact basis functions decomposition as well as measurement noise. Third, we illustrate our results on a model of a fully-actuated double inverted pendulum. An accompanying technical report can be found in \cite{Alsalti2022b}, which provides the detailed theoretical analysis of recursive feasibility and practical exponential stability of the proposed scheme, i.e., the proof of Theorem~\ref{rob_npc_thm} below.
	\section{Preliminaries}\label{prel}
	\subsection{Notation}
	The set of integers in the interval $[a,b]$ is denoted by $\mathbb{Z}_{[a,b]}$. For a vector $\mu\in\mathbb{R}^n$ and a positive definite symmetric matrix $P=P^\top\succ0$, the $p-$norm is given by $\norm*{\mu}_p$ for $p=1,2,\infty$, whereas $\norm*{\mu}_P = \sqrt{\mu^\top P\mu}$. The minimum and maximum singular values of the matrix $P$ are denoted by $\sigma_{\textup{min}}(P),\sigma_{\textup{max}}(P)$, respectively. The induced norm of a matrix $P$ is denoted by $\norm*{P}_i$ for $i=1,2,\infty$. We use $\mathbf{0}$ to denote a vector or matrix of zeros of appropriate dimensions.	For a sequence $\{\mathbf{z}_k\}_{k=0}^{N-1}$ with $\mathbf{z}_k\in\mathbb{R}^\eta$, each element is expressed as $\mathbf{z}_k=\begin{bmatrix}
		z_{1,k} & z_{2,k} & \dots & z_{\eta,k}
	\end{bmatrix}^\top$. The stacked vector of that sequence is given by $\mathbf{z}=\begin{bmatrix}
		\mathbf{z}_0^\top & \dots & \mathbf{z}_{N-1}^\top
	\end{bmatrix}^\top$, and a window of it by $\mathbf{z}_{[a,b]}=\begin{bmatrix}
		\mathbf{z}_a^\top & \dots & \mathbf{z}_b^\top
	\end{bmatrix}^\top$. The Hankel matrix of depth $L$ of this sequence is given by
	\begin{equation*}
		\begin{aligned}
			H_L(\mathbf{z})&=\begin{bmatrix}
				\mathbf{z}_0 & \mathbf{z}_1 & \dots & \mathbf{z}_{N-L}\\
				\mathbf{z}_1 & \mathbf{z}_2 & \dots & \mathbf{z}_{N-L+1}\\
				\vdots & \vdots & \ddots & \vdots\\
				\mathbf{z}_{L-1} & \mathbf{z}_L & \dots & \mathbf{z}_{N-1}
			\end{bmatrix}.
		\end{aligned}
	\end{equation*}
	Throughout the paper, the notion of persistency of excitation (PE) is defined as follows.
	\begin{definition}
		The sequence $\{\mathbf{z}_k\}_{k=0}^{N-1}$ is said to be persistently exciting of order $L$ if \textup{rank}$\left(H_L(\mathbf{z})\right)=\eta L$.
	\end{definition}
	\subsection{Willems' fundamental lemma}
	In the following, we recall the main result of \cite{Willems05} in the state-space framework. Consider the following DT-LTI system of the form
	\begin{equation}
		\begin{matrix}
			\mathbf{x}_{k+1}=A\mathbf{x}_k+B\mathbf{u}_k,&\quad&\mathbf{y}_k=C\mathbf{x}_k+D\mathbf{u}_k
		\end{matrix}
		\label{DTLTI}
	\end{equation}
	where \(\mathbf{x}_k\in\mathbb{R}^n\) is the state, \(\mathbf{u}_k\in\mathbb{R}^m\) is the input, \(\mathbf{y}_k\in\mathbb{R}^p\) is the output and the pair \((A,B)\) is controllable. A sequence $\{\mathbf{u}_k,\mathbf{y}_k\}_{k=0}^{N-1}$ is said to be an input-output trajectory of \eqref{DTLTI} if there exists an initial condition $\mathbf{x}_0$ such that \eqref{DTLTI} holds for all $k\in\mathbb{Z}_{[0,N-1]}$. The \textit{fundamental lemma} is stated as follows.\par
	\begin{thm}[\hspace{-0.25mm}\cite{Willems05}]\label{WFL}
		Let \(\{\mathbf{u}_k,\mathbf{y}_k\}_{k=0}^{N-1}\) be an input-output trajectory of the controllable system \eqref{DTLTI}. If \(\{\mathbf{u}_k\}_{k=0}^{N-1}\) is persistently exciting of order \(L+n\), then
		\begin{itemize}
			\item[(i)] the matrix $\begin{bmatrix}
				H_L(\mathbf{u}) \\ H_1(\mathbf{x}_{[0,N-L]})
			\end{bmatrix}$ has full row rank, and
			\item[(ii)] any \(\{\bar{\mathbf{u}}_k,\bar{\mathbf{y}}_k\}_{k=0}^{L-1}\) is an input-output trajectory of \eqref{DTLTI} if and only if there exists \(\alpha\in\mathbb{R}^{N-L+1}\) such that
			\begin{equation}
				\begin{bmatrix} H_L(\mathbf{u})\\ H_L(\mathbf{y})\end{bmatrix} \alpha = \begin{bmatrix}\bar{\mathbf{u}} \\ \bar{\mathbf{y}}\end{bmatrix}.
				\label{fundamental_lemma}
			\end{equation}
		\end{itemize}
	\end{thm}
\subsection{Discrete-time feedback linearizable systems}\label{sec_MBFLsys}
Consider the following DT-MIMO square nonlinear system
\begin{equation}
	\begin{matrix}
		\mathbf{x}_{k+1} = \boldsymbol{f}(\mathbf{x}_k,\mathbf{u}_k), &\quad& \mathbf{y}_k = \boldsymbol{h}(\mathbf{x}_k),
	\end{matrix}
\label{NLsys}
\end{equation}
where $\mathbf{x}_k\in\mathbb{R}^n$ is the state vector and $\mathbf{u}_k,\mathbf{y}_k\in\mathbb{R}^m$ are the input and output vectors, respectively. The functions $\boldsymbol{f}:\mathbb{R}^n\times\mathbb{R}^m\to\mathbb{R}^n$, $\boldsymbol{h}:\mathbb{R}^n\to\mathbb{R}^m$ are analytic functions with $\boldsymbol{f}(\mathbf{0},\mathbf{0})=\mathbf{0}$ and $\boldsymbol{h}(\mathbf{0})=\mathbf{0}$.\par
As defined in \cite{MonacoNor1987}, each output $y_i=h_i(\mathbf{x})$ of the nonlinear system \eqref{NLsys}, for $i\in\mathbb{Z}_{[1,m]}$, is said to have a (globally) well-defined relative degree $d_i$ for all $\mathbf{x}_k\in\mathbb{R}^n$ and $\mathbf{u}_k\in\mathbb{R}^m$ if at least one of the $m$ inputs at time $k$ affects the $i-$th output at time $k+d_i$. In particular,
\begin{equation}
\begin{aligned}
	y_{i,k+d_i} &= h_i(\boldsymbol{f}_O^{d_i-1}(\boldsymbol{f}(\mathbf{x}_k,\mathbf{u}_k))),
	\label{output_k}
\end{aligned}
\end{equation}
where $\boldsymbol{f}_O^{j}$ is the $j-$th iterated composition of the undriven dynamics $f(\mathbf{x}_k,\mathbf{0})$. In the remainder of the paper, we denote the maximum relative degree by $d_{\textup{max}}\coloneqq\max_i d_i$. Note that $d_{\textup{max}}$ is also the system's controllability index.\par
For a system to be full-state feedback linearizable, the sum of relative degrees must equal $n$, i.e., $\sum_i d_i=n$. For globally well-defined relative degrees, this condition can be checked by perturbing the system from rest and recording the first time instants at which each output changes from zero. If the sum of all these instances is $n$, then the system is full-state feedback linearizable. In this case, one can show, under standard assumptions, that the system \eqref{NLsys} can be transformed into DT normal form (compare, e.g., \cite{MonacoNor1987, Alsalti2022a}). This means that there exists an invertible (w.r.t. $\mathbf{v}_k$) control law $\mathbf{u}_k=\gamma(\mathbf{x}_k,\mathbf{v}_k)$, with $\gamma:\mathbb{R}^n\times\mathbb{R}^m\to\mathbb{R}^m$ and an invertible coordinate transformation $\Xi_k=T(\mathbf{x}_k)$, such that the system can be written as
\begin{equation}
	\begin{matrix}
		\Xi_{k+1} = \mathcal{A}\Xi_k + \mathcal{B}\mathbf{v}_k,&\quad
		\mathbf{y}_k = \mathcal{C}\Xi_k,
	\end{matrix}
	\label{BINF}%
\end{equation}%
where $\Xi_k\in\mathbb{R}^{n}$ is defined as
\begin{equation}
	\begin{aligned}
		\Xi_k &=\begin{bmatrix}
			y_{1,[k,k+d_1-1]}^\top & \dots & y_{m,[k,k+d_m-1]}^\top
		\end{bmatrix}^\top.
	\end{aligned}
	\label{Xi}
\end{equation}
Further, $\mathcal{A},\,\mathcal{B},\,\mathcal{C}$ are in block-Brunovsky\footnote{See \cite[Appendix A]{Alsalti2022a} for the structure of the block-Brunovsky form.} form, which is a controllable/observable triplet.
\subsection{Willems' lemma for feedback linearizable systems}
In \cite{Alsalti2022a}, the fundamental lemma \citep{Willems05} was extended to the class of feedback linearizable systems as in \eqref{NLsys}. This was done by first noticing that \eqref{BINF} is a linear system in transformed coordinates. It follows from Theorem \ref{WFL} that if $\{\mathbf{v}_k\}_{k=0}^{N-1}$ is persistently exciting of order $L+n$, then any $\{\mathbf{\bar{v}}_k,\mathbf{\bar{y}}_k\}_{k=0}^{L-1}$ is a trajectory of \eqref{BINF} if and only if there exists $\alpha\in\mathbb{R}^{N-L+1}$ such that $\begin{bsmallmatrix}
		H_L(\mathbf{v})\\ H_L(\mathbf{y})
	\end{bsmallmatrix}\alpha=\begin{bsmallmatrix}
		\bar{\mathbf{v}}\\[0.75mm] \bar{\mathbf{y}}
	\end{bsmallmatrix}$ holds. However, one typically only has access to input-output data (i.e., $\mathbf{u}_k,\mathbf{y}_k$) and not to the synthetic input $\mathbf{v}_k$ or to the corresponding state transformation $\Xi_k = T(\mathbf{x}_k)$.\par
In order to come up with a data-based description of the trajectories of the feedback linearizable nonlinear system \eqref{NLsys}, the synthetic input can be expressed as\footnote{It was shown in \cite{MonacoNor1987} that $\tilde{\Phi}$ is in fact an iterated composition of the analytic functions $\boldsymbol{f},\boldsymbol{h}$ in \eqref{NLsys} and, hence, it is locally Lipschitz continuous. By local Lipschitz continuity of $T^{-1}$ (cf. \eqref{output_k} and \eqref{Xi}), $\Phi$ is locally Lipschitz continuous as well.}
\begin{equation*}
	\begin{aligned}
		\mathbf{v}_k = \tilde{\Phi}(\mathbf{u}_k,\mathbf{x}_k) = \tilde{\Phi}(\mathbf{u}_k,T^{-1}(\Xi_k))\eqqcolon \Phi(\mathbf{u}_k,\Xi_k).
	\end{aligned}
\end{equation*}
This allows us to parameterize $\mathbf{v}_k$ using input-output data only since $\Xi_k$ is given by shifted output data (see \eqref{Xi}). Since the function $\Phi$ is unknown, a user-defined dictionary of basis functions, which only depend on input-output data, is used to approximate it. In particular,
\begin{align}
		\Phi(\mathbf{u}_k,\Xi_k) &\eqqcolon \mathcal{G}\Psi(\mathbf{u}_k,\Xi_k) + \scalebox{1.5}{$\epsilon$}(\mathbf{u}_k,\Xi_k)		\label{basis}\\
		\begin{bmatrix}
			\phi_1(\mathbf{u}_k,\Xi_k)\\ \vdots \\ \phi_m(\mathbf{u}_k,\Xi_k)
		\end{bmatrix}&= \begin{bmatrix}
			\rule[.5ex]{1.0em}{0.4pt} \,g_1^\top\, \rule[.5ex]{1.0em}{0.4pt}\\
			\vdots\\
			\rule[.5ex]{1.0em}{0.4pt} \,g_m^\top\, \rule[.5ex]{1.0em}{0.4pt}\\
		\end{bmatrix}\Psi(\mathbf{u}_k,\Xi_k) + \begin{bmatrix}
			\varepsilon_1(\mathbf{u}_k,\Xi_k)\\ \vdots\\ \varepsilon_m(\mathbf{u}_k,\Xi_k)
		\end{bmatrix},\notag
\end{align}
where $\Psi(\mathbf{u}_k,\Xi_k)$ is the vector of $r\in\mathbb{N}$ locally Lipschitz continuous and linearly independent basis functions $\psi_j~:~\mathbb{R}^m\times\mathbb{R}^n\to\mathbb{R}$, for $j\in\mathbb{Z}_{[1,r]}$, and $\scalebox{1.5}{$\epsilon$}(\mathbf{u}_k,\Xi_k)$ is the vector of approximation errors $\varepsilon_i:\mathbb{R}^m\times\mathbb{R}^n\to\mathbb{R}$ for $i\in\mathbb{Z}_{[1,m]}$. The matrix $\mathcal{G}\in\mathbb{R}^{m\times r}$, whose rows are $g_i^\top$ for all $i\in\mathbb{Z}_{[1,m]}$, is the matrix of unknown coefficients that is assumed to have full row rank and, hence, has a right inverse $\mathcal{G}^\dagger\coloneqq \mathcal{G}^\top(\mathcal{GG}^\top)^{-1}$. For the following analysis, we define the matrix $\mathcal{G}$ as
\begin{equation}
	\mathcal{G}\coloneqq\argmin\limits_{{G}}\left\langle \Phi- {G}\Psi, \Phi - {G}\Psi \right\rangle,\label{def_of_G_mat}
\end{equation}
where the inner product on a compact subset of the input-state space\footnote{Compare Section \ref{robust_nmpc_sec} and Assumption \ref{bounded_err_assmp} below for a discussion regarding the set $\Omega$.} $\Omega\subset\mathbb{R}^m\times\mathbb{R}^n$ is given by
\begin{equation*}
	\left< \rho_1, \rho_2\right> = \int_\Omega \rho_1(\ell_1,\ell_2)\rho_2(\ell_1,\ell_2) d\ell_1 d\ell_2.
\end{equation*}
We highlight that the matrix $\mathcal{G}$ in \eqref{def_of_G_mat} is never computed in this work. Computing $\mathcal{G}$ amounts to identifying a model of the system, which is not the focus of this work.\par
In practice, the measured output data is typically noisy. In what follows, we denote it by $\mathbf{\tilde{y}}_k=\mathbf{y}_k+\mathbf{w}_k$, and assume that $\norm*{\mathbf{w}_k}_\infty\leq {w}^*$, for all $k\geq0$, is a uniformly bounded output measurement noise. Using noisy measurements in place of the noiseless output data on the right hand side of \eqref{basis} results in an additional error term. In particular,
\begin{equation}
	\Phi(\mathbf{u}_k,\Xi_k) = \mathcal{G}\Psi(\mathbf{u}_k,\tilde{\Xi}_k) + \scalebox{1.5}{$\epsilon$}(\mathbf{u}_k,\tilde{\Xi}_k) + \delta(\omega_k),
	\label{noisy_basis}
\end{equation}
where $\tilde{\Xi}_k = \Xi_k+\omega_k$, $\omega_k=\begin{bmatrix}w_{1,[k,k+d_1-1]}^\top\, \cdots\, w_{m,[k,k+d_m-1]}^\top\end{bmatrix}^\top$ and $\delta(\omega_k)\coloneqq \mathcal{G}\Psi(\mathbf{u}_k,{\Xi}_k) + \scalebox{1.5}{$\epsilon$}(\mathbf{u}_k,{\Xi}_k) - \mathcal{G}\Psi(\mathbf{u}_k,\tilde{\Xi}_k) - \scalebox{1.5}{$\epsilon$}(\mathbf{u}_k,\tilde{\Xi}_k)$. Substituting back into \eqref{BINF}, we get
\begin{align}
	\Xi_{k+1} &= \mathcal{A}\Xi_k + \mathcal{B}\mathcal{G}(\hat{\Psi}_k(\mathbf{u},{\tilde{\Xi}})+\hat{E}_k(\mathbf{u},{\tilde{\Xi}}) + \hat{D}_k(\omega)),\notag\\
	{\mathbf{\tilde{y}}_k }&{= \mathcal{C}\Xi_k + \mathbf{w}_k},	\label{Lsys3}
\end{align}
where
\begin{gather}
		\hat{\Psi}_k(\mathbf{u},{\tilde{\Xi}}) \coloneqq \Psi(\mathbf{u}_k,{\tilde{\Xi}_k}), \quad \hat{D}_k(\omega) \coloneqq D(\omega_k)= \mathcal{G}^\dagger\delta(\omega_k),\notag\\
		\hat{E}_k(\mathbf{u},{\tilde{\Xi}}) \coloneqq E(\mathbf{u}_k,{\tilde{\Xi}_k})= \mathcal{G}^\dagger\scalebox{1.5}{$\epsilon$}(\mathbf{u}_k,{\tilde{\Xi}_k}).
\label{imp_defs}
\end{gather}
Similarly, we use the following notation
\begin{equation*}
	\begin{matrix}
		\hat{\delta}_k(\omega) \coloneqq \delta(\omega_k), \quad \hat{\varepsilon}_{i,k}(\mathbf{u},{\tilde{\Xi}}) \coloneqq 	\varepsilon_i(\mathbf{u}_k,{\tilde{\Xi}_k}).
	\end{matrix}
\end{equation*}
Due to the block-Brunovksy form and $\mathcal{G}$ having full row rank by assumption, the pair $(\mathcal{A,BG})$ in \eqref{Lsys3} is a controllable pair \cite[Lemma 1]{Alsalti2022a}. Further, by the structure of the system \eqref{Lsys3}, the $i-$th synthetic input at time $k$ only affects (in fact, is equal to) the $i-$th \textit{noiseless} output at time $k+d_i$, i.e.,
\begin{align}
		\hspace{-1mm}y_{i,k+d_i}\hspace{-1mm} \stackrel{\eqref{BINF}}{=}\hspace{-1mm} v_{i,k} \hspace{-1mm}&=\hspace{-1mm} \phi_i(\mathbf{u}_k,\Xi_k)\hspace{-1.5mm}\stackrel{\eqref{basis}, \eqref{imp_defs}}{=}\hspace{-1.5mm} g_i^\top\hspace{-0.5mm}(\hat{\Psi}_k(\mathbf{u},{{\Xi}})+\hat{E}_k(\mathbf{u},{{\Xi}}))\hspace{-1mm}\notag\\
		&\stackrel{\eqref{noisy_basis}}{=}g_i^\top(\hat{\Psi}_k(\mathbf{u},{\tilde{\Xi}})+\hat{E}_k(\mathbf{u},{\tilde{\Xi}}) + \hat{D}_k(\omega)).\label{y_for_proofs}
\end{align}
The following theorem provides an input-output data-based representation of \eqref{BINF} in the nominal case, i.e., when $\hat{E}_k(\mathbf{u},\tilde{\Xi})\equiv\hat{D}_k(\omega)\equiv0$. In Section \ref{DBNPC}, this  representation will serve as a predictive model in the data-driven nonlinear predictive control scheme.
\begin{thm}\cite[Thm. 2]{Alsalti2022a}\label{DB_MIMO_rep}
	Let $\{\mathbf{u}_k^{\textup{d}}\}_{k=0}^{N-1}$, $\{y_{i,k}^{\textup{d}}\}_{k=0}^{N+d_i-1}$, for $i\in\mathbb{Z}_{[1,m]}$, be a trajectory of a full-state feedback linearizable system as in \eqref{NLsys}. Furthermore, let $\{\hat{\Psi}_k(\mathbf{u}^{\textup{d}},\Xi^{\textup{d}})\}_{k=0}^{N-1}$ be persistently exciting of order $L+n$. Then, any $\{\mathbf{\bar{u}}_k\}_{k=0}^{L-1}$, $\{\bar{y}_{i,k}\}_{k=0}^{L+d_i-1}$ is a trajectory of system \eqref{NLsys} if and only if there exists $\alpha\in\mathbb{R}^{N-L+1}$ such that
	\begin{equation}
		\begin{bmatrix}
			H_{L}(\hat{\Psi}(\mathbf{u}^{\textup{d}},\Xi^{\textup{d}}))\\ 	H_{L+1}(\Xi^{\textup{d}})\end{bmatrix}\alpha = \begin{bmatrix}
			\hat{\Psi}(\mathbf{\bar{u}},\bar{\Xi})\\ 	\bar{\Xi}
		\end{bmatrix},
		\label{inexact_result}
	\end{equation}
	where $\hat{\Psi}(\mathbf{\bar{u}},\bar{\Xi})$ is the stacked vector of the sequence $\{\hat{\Psi}_k(\mathbf{\bar{u}},\bar{\Xi})\}_{k=0}^{L-1}$, while $\Xi^{\textup{d}},\,\bar{\Xi}$ are the stacked vectors of $\{\Xi_k^{\textup{d}}\}_{k=0}^{N},\,\{\bar{\Xi}_k\}_{k=0}^{L}$ which, as in \eqref{Xi}, are composed of $\{y_{i,k}^{\textup{d}}\}_{k=0}^{N+d_i-1},\,\{\bar{y}_{i,k}\}_{k=0}^{L+d_i-1}$, respectively.
\end{thm}
	\section{Data-driven nonlinear predictive control}\label{DBNPC}
In this section, we present and analyze the data-driven nonlinear predictive control scheme. In Section \ref{nom_scheme}, we present the nominal data-based nonlinear predictive control scheme, where the basis function approximation error is zero for all time and the output data is noiseless. In Section \ref{robust_nmpc_sec}, we present the more realistic case when the error terms in \eqref{noisy_basis} are nonzero but uniformly upper bounded.
\subsection{Nominal scheme}
\label{nom_scheme}
For the case when $\hat{E}_k(\mathbf{u},\tilde{\Xi})\equiv\hat{D}_k(\omega)\equiv0$, Theorem \ref{DB_MIMO_rep} provides an exact data-based representation of all trajectories of System \eqref{NLsys}. This representation will serve as the prediction model in the predictive control scheme. Specifically, given a prediction horizon $L\geq0$, the following nonlinear minimization problem is solved at each $t\in\mathbb{N}$ in a receding horizon fashion
\begin{subequations}
	\begin{align}
		J^*_t =&\min\limits_{\tiny\begin{matrix}\mathbf{\bar{u}}(t),\bar{y}_i(t)\\\alpha(t)\end{matrix}}\;\sum\limits_{k=0}^{L-1}\ell(\mathbf{\bar{u}}_k(t),\mathbf{\bar{y}}_k(t))\label{stage_costs_nom_nmpc}\\
		\text{s.t.}\;&\begin{bmatrix}\hat{\Psi}(\mathbf{\bar{u}}(t),\bar{\Xi}(t))\\ \bar{\Xi}(t)\end{bmatrix}\hspace{-1.5mm} =\hspace{-1.5mm}
		\begin{bmatrix}\hspace{-0.75mm} H_{L+d_{\textup{max}}}(\hat{\Psi}(\mathbf{u}^{\textup{d}},{\Xi}^{\textup{d}})) \hspace{-0.75mm}\\ H_{L+d_{\textup{max}}+1}(\Xi^{\textup{d}})\end{bmatrix}\hspace{-1mm} \alpha(t),\label{pc_willems}\\
		&\begin{bmatrix}
			\mathbf{\bar{u}}_{[-d_{\textup{max}},-1]}(t)\\
			\mathbf{\bar{y}}_{[-d_{\textup{max}},-1]}(t)
		\end{bmatrix} = \begin{bmatrix}
			\mathbf{{u}}_{[t-d_{\textup{max}},t-1]}\\
			\mathbf{{y}}_{[t-d_{\textup{max}},t-1]}
		\end{bmatrix}\label{pc_ini},\\
		&\bar{y}_{i,[L,L+d_i-1]}(t) = \mathbf{0},\quad \forall i\in\mathbb{Z}_{[1,m]},\label{pc_term}\\
		&\bar{\Xi}_k(t)\hspace{-1mm}=\hspace{-1mm}\begin{bmatrix}
			\bar{y}_{1,[k,k+d_1-1]}^\top(t) & \dots & \bar{y}_{m,[k,k+d_m-1]}^\top(t)
		\end{bmatrix}^{\hspace{-1mm}\top}\hspace{-2mm},\label{pc_fixing_Xi}\\
		&\mathbf{\bar{u}}_k(t)\in\mathcal{U},
			\mathbf{\bar{y}}_k(t)\in\mathcal{Y},\quad\forall k\in\mathbb{Z}_{[-d_{\textup{max}},L-1]}.\label{CL_constraints}
	\end{align}%
	\label{DDNMPC}%
\end{subequations}
The notation used in \eqref{DDNMPC} is summarized as follows. The sequences $\{\mathbf{u}_k^{\textup{d}}\}_{k=0}^{N-1},\{\Xi_k^{\textup{d}}\}_{k=0}^{N}$ denote previously collected input-state data (where the latter is constructed from shifted outputs $\{y_{i,k}^{\textup{d}}\}_{k=0}^{N+d_i-1}$, for $i\in\mathbb{Z}_{[1,m]}$, as in \eqref{Xi}). The sequences $\{\mathbf{\bar{u}}_k(t)\}_{k=-d_{\textup{max}}}^{L-1},\{\bar{\Xi}_k(t)\}_{k=-d_{\textup{max}}}^{L}$ and $\{{\bar{y}}_{i,k}(t)\}_{k=-d_{\textup{max}}}^{L+d_i-1}$ denote the predicted input, state and output trajectories, predicted at time $t$, whereas $\mathbf{u}_t,\Xi_t,y_{i,t}$ denote the current input, state, and outputs of the system at time $t$, respectively. Notice that $\{\bar{\Xi}_k(t)\}_{k=-d_{\textup{max}}}^{L}$ and $\{{\bar{y}}_{i,k}(t)\}_{k=-d_{\textup{max}}}^{L+d_i-1}$ are related by \eqref{pc_fixing_Xi}. The optimal predicted input, state and output sequences are denoted by $\{\mathbf{\bar{u}}_k^*(t)\}_{k=-d_{\textup{max}}}^{L-1},\{\bar{\Xi}_k^*(t)\}_{k=-d_{\textup{max}}}^{L}$ and $\{{\bar{y}}_{i,k}^*(t)\}_{k=-d_{\textup{max}}}^{L+d_i-1}$. In \eqref{stage_costs_nom_nmpc}, we consider quadratic stage cost functions that penalize the difference of the predicted input and output trajectories from a desired equilibrium that is known a priori. In particular, we consider
\begin{equation}
	\ell(\mathbf{\bar{u}}_k(t),\mathbf{\bar{y}}_k(t)) = \norm*{\mathbf{\bar{u}}_k - \mathbf{u}^s}_R^2 + \norm*{\mathbf{\bar{y}}_k - \mathbf{y}^s}_Q^2,\label{stage_cost_fcn}
\end{equation}
where $Q = Q^\top \succ0$, $R=R^\top \succ 0$ and $\mathbf{u}^s,\,\mathbf{y}^s$ are the input and output equilibrium points corresponding to an equilibrium state $\mathbf{x}^s$ for the system in \eqref{NLsys}. In the following theoretical analysis, we consider deviations from the origin, i.e., $\mathbf{u}^s=\mathbf{y}^s=\mathbf{0}$. The results similarly apply to non-zero fixed equilibrium points subject to changes in the structure of some constants of the proofs.\par
Note that \eqref{pc_willems} uses \eqref{inexact_result} (but for longer sequences) to generate the predicted inputs and outputs $\mathbf{\bar{u}}(t),\bar{y}_i(t)$ (cf. Theorem \ref{DB_MIMO_rep}). The increased length of the sequences is caused by the need to fix the state at time $t-d_{\textup{max}}$ using the past $d_{\textup{max}}$ instances of the output (see \eqref{pc_ini}). In contrast, the past $d_{\textup{max}}$ instances of the input are used to \emph{implicitly} fix the state at time $t$, through their effect on the outputs $y_{i,[t,t+d_{\textup{max}}-1]}$ (see \eqref{y_for_proofs}). To show asymptotic stability of the proposed predictive control scheme, we follow standard arguments from the MPC literature \cite{Rawlings17}. Specifically, we use terminal equality constraints \eqref{pc_term} that force the predicted state to be zero at the end of the prediction horizon. Finally, the system's inputs and outputs are subject to pointwise-in-time constraints, i.e., $\mathbf{{u}}_t\in\mathcal{U}\subseteq\mathbb{R}^m$, $\mathbf{y}_t\in\mathcal{Y}\subseteq\mathbb{R}^m$, and we assume that $(\mathbf{u}^s,\mathbf{y}^s)\in\text{int}(\mathcal{U}\times\mathcal{Y})$.\par
Before stating the main of this subsection result, we make the following assumptions.
\begin{assumption}\label{J_ub_asm}
	The optimal value function $J^*_t$ is upper bounded by a class $\mathcal{K}$ function \citep{Kellet14}, i.e., $J^*_t \leq \beta_1(\norm{\Xi_t})$ for all $\Xi_t\in\mathbb{X}\coloneqq\{\Xi\in\mathbb{R}^n~|~J^*_t<\infty\}$.
\end{assumption}
\begin{assumption}\label{PE_asmp}
	The input-output data are collected such that $\hat{\Psi}(\mathbf{u}^{\textup{d}},{\Xi}^{\textup{d}})$ is persistently exciting of order $L+d_{\textup{max}}+n$.
\end{assumption}
Assumption \ref{J_ub_asm} is similar to standard assumptions made in MPC (cf. \cite{Rawlings17}, Assumption 2.17). Assumption \ref{PE_asmp} is needed to apply Theorem \ref{DB_MIMO_rep}. This condition can be checked after collecting the offline data. Alternatively, in \cite{Alsalti2023}, a method to design inputs resulting in persistently exciting sequences of basis functions is proposed.
Notice that persistency of excitation is of order $L+d_{\textup{max}}+n$ unlike Theorem \ref{DB_MIMO_rep} where the order was $L+n$. This is because the prediction horizon is extended to include the previous $d_{\textup{max}}$ instances of the input and output (see \eqref{pc_ini}).\par
The following result shows that the nominal nonlinear predictive-control scheme in \eqref{DDNMPC} is recursively feasible, satisfies the constraints and that the resulting closed-loop system is asymptotically stable.
\begin{thm}\label{nom_NPC_thm}
	Suppose Assumption~\ref{J_ub_asm} and \ref{PE_asmp} are satisfied. If \eqref{DDNMPC} is feasible at $t=0$, then
	\begin{itemize}
		\item[(i)] it is recursively feasible for all $t\in\mathbb{N}$,
		\item[(ii)] the closed-loop satisfies the constraints \eqref{CL_constraints}, and
		\item[(iii)] the equilibrium $\mathbf{x}^s=0$ is asymptotically stable.
	\end{itemize}
\end{thm}
\begin{pf}
	The proof follows similar arguments as in \cite{Berberich203} for linear systems. First, recursive feasibility and constraint satisfaction can be shown using a candidate solution defined by shifting the previously optimal solution and appending it by zero. Asymptotic stability is shown using standard Lyapunov arguments along with IOSS properties. However, unlike \cite{Berberich203} the optimal value function $J^*_t$ (and hence the Lyapunov function candidate) is upper bounded by class $\mathcal{K}$ function (cf. Assumption \ref{J_ub_asm}) and not necessarily by a quadratic function in the state $\Xi_t$, due to the nonlinear constraints \eqref{pc_willems}. This results in asymptotic stability (cf. \cite[Thm. 4.16]{Khalil2002}) in contrast to the exponential stability of the LTI setting.\hfill$\blacksquare$
\end{pf}
\subsection{Robust scheme}
\label{robust_nmpc_sec}
Now we consider the practically relevant case when the basis functions approximation error is non-zero and unknown {and the output data is noisy}. To account for the nonzero error terms in \eqref{noisy_basis}, we consider a compact subset of the input-state space, i.e., $\Omega\subset\mathbb{R}^m\times\mathbb{R}^n$, in which the trajectories of the system (offline and online) evolve\footnote{We prove in Theorem \ref{rob_npc_thm} that the closed-loop trajectories indeed evolve in a compact subset of the input-state space.}. This, along with local Lipschitz continuity of $\Phi$ and $\Psi$, guarantees a uniform upper bound on the approximation error. To this end, we make the following assumption.
\begin{assumption}\label{bounded_err_assmp}
	The error in the basis function approximation $\hat{\scalebox{1.5}{$\epsilon$}}_k(\mathbf{u},{{\Xi}})$ is uniformly upper bounded by a known $\varepsilon^*>0$, i.e., $\norm*{\hat{\scalebox{1.5}{$\epsilon$}}_k(\mathbf{u},{{\Xi}})}_\infty\leq\varepsilon^*$, for all $(\mathbf{u}_k,{\Xi}_k)\in\Omega\subset\mathbb{R}^m\times\mathbb{R}^n$, where $\Omega$ is a compact subset of the input-state space.
\end{assumption}
\begin{remark}
	Recall that $\hat{E}_k(\mathbf{u},\Xi)\stackrel{\eqref{imp_defs}}{=} \mathcal{G}^\dagger\hat{\scalebox{1.5}{$\epsilon$}}_k(\mathbf{u},\Xi)$, and hence, Assumption \ref{bounded_err_assmp} implies
	\begin{equation}
		\begin{aligned}
			\norm{\hat{E}_k(\mathbf{u},\Xi)}_\infty &{\leq} \norm{\mathcal{G}^\dagger}_\infty\norm{\hat{\scalebox{1.5}{$\epsilon$}}_k(\mathbf{u},\Xi)}_\infty\leq\norm{\mathcal{G}^\dagger}_\infty\varepsilon^*.
		\end{aligned}\label{tilde_E_bdd_asmp}
	\end{equation}
\end{remark}
\begin{remark}\label{remark_Kw}
	Note that $\delta(\omega_k)$ in \eqref{noisy_basis} satisfies $\delta(\mathbf{0})=\mathbf{0}$ and, by local Lipschitz continuity of $\Phi$ and $\Psi$ on a compact set $\Omega$ as well as boundedness of $\mathbf{w}_k$, there exists a $K_w>0$ such that $\norm*{\delta(\omega_k)}_\infty\leq K_w w^*$ for all $k\geq0$.
\end{remark}
\begin{remark}\label{remark_KPsiKXi}
	The Lipschitz constants of $\Psi$ and $\Phi$ w.r.t. their second argument on the compact set $\Omega$ are denoted by $K_\Psi$ and $K_\Xi$, respectively.
\end{remark}
The following optimization problem provides the basis for the proposed robust predictive control approach.
\begin{subequations}
	\begin{align}
		J^*_t&=\min\limits_{\tiny\begin{matrix}\mathbf{\bar{u}}(t),\bar{y}_i(t)\\\alpha(t),\sigma(t)\end{matrix}}\hspace{-0.5mm}\sum\limits_{k=0}^{L-1}\hspace{-0.5mm}\ell(\mathbf{\bar{u}}_k(t),\mathbf{\bar{y}}_k(t))\hspace{-0.5mm}+\hspace{-0.75mm}\lambda_\alpha\hspace{-0.25mm}\max\{\varepsilon^*,w^*\}\norm*{\alpha(t)}_2^2\hspace{-0.5mm}\notag\\[-2.5ex]
		&\qquad\qquad\qquad+\hspace{-0.5mm}\lambda_\sigma\norm*{\sigma(t)}_2^2\label{opt_val_fcn_def}\\[1ex]
		\text{s.t.}&\begin{bmatrix}\hspace{-0.5mm}\hat{\Psi}(\mathbf{\bar{u}}(t),\bar{\Xi}(t))\hspace{-0.5mm} \\ \bar{\Xi}(t)\end{bmatrix}\hspace{-1.75mm}+\hspace{-1mm}\sigma(t)\hspace{-1mm} =\hspace{-1.5mm}
		\begin{bmatrix}\hspace{-0.5mm} H_{L+d_{\textup{max}}}\hspace{-0.5mm}(\hspace{-0.5mm}\hat{\Psi}(\mathbf{u}^{\textup{d}}\hspace{-0.5mm},{\tilde{\Xi}^{\textup{d}}}\hspace{-0.5mm})\hspace{-0.5mm}) \hspace{-0.5mm} \\ H_{L+d_{\textup{max}}+1}({\tilde{\Xi}^{\textup{d}}})\end{bmatrix}\hspace{-1mm} \alpha(t),\label{pc2_willems}\\
		&\begin{bmatrix}
			\mathbf{\bar{u}}_{[-d_{\textup{max}},-1]}(t)\\
			\mathbf{\bar{y}}_{[-d_{\textup{max}},-1]}(t)
		\end{bmatrix} = \begin{bmatrix}
			\mathbf{{u}}_{[t-d_{\textup{max}},t-1]}\\
			{\mathbf{\tilde{y}}_{[t-d_{\textup{max}},t-1]}}
		\end{bmatrix}\label{pc2_ini},\\
		&\bar{y}_{i,[L,L+d_i-1]}(t) = \mathbf{0},\quad \forall i\in\mathbb{Z}_{[1,m]},\label{pc2_term}\\
		&\bar{\Xi}_k(t)\hspace{-1mm}=\hspace{-1mm}\begin{bmatrix}
			\bar{y}_{1,[k,k+d_1-1]}^\top(t) & \dots & \bar{y}_{m,[k,k+d_m-1]}^\top(t)
		\end{bmatrix}^{\hspace{-1mm}\top}\hspace{-2mm},\label{pc2_fixing_Xi}\\
		&\norm*{\hspace{-0.25mm}\sigma_k(t)\hspace{-0.25mm}}_{\hspace{-0.5mm}\infty}\hspace{-1mm}\leq \hspace{-1mm}K_{\Psi}w^* \hspace{-1mm}+\hspace{-1mm} (\hspace{-0.25mm}\varepsilon^* \hspace{-1mm}+ \hspace{-1mm}K_w w^*\hspace{-0.5mm})\norm*{\mathcal{G}^\dagger}_{\hspace{-0.5mm}\infty}\hspace{-1mm}(1\hspace{-1mm}+\hspace{-1mm}\norm*{\alpha(t)}_1\hspace{-0.5mm})\hspace{-0.5mm},\label{slack_const}\\
		&\mathbf{\bar{u}}_k(t)\in\mathcal{U},\quad\forall k\in\mathbb{Z}_{[-d_{\textup{max}},L-1]}.\label{rob_input_const_set}
	\end{align}%
	\label{DDNMPC2}%
\end{subequations}
The presence of nonzero error terms in \eqref{noisy_basis}, in general, makes it difficult to satisfy \eqref{pc_willems} since Assumption \ref{PE_asmp} only guarantees existence of solutions $\alpha(t)$ that satisfy \eqref{pc_willems} in case of noiseless data. To account for this, we relax \eqref{pc_willems} by introducing a slack variable $\sigma(t) = [\sigma_{\Psi}^\top(t)\quad \sigma_{\Xi}^\top(t)]^\top$ as in \eqref{pc2_willems} and penalize its $2-$norm in the cost function (compare \cite{Coulson20, Berberich203}). Additionally, we require the slack variable to be bounded as in \eqref{slack_const}. Furthermore, to mitigate the effect of uncertainties on the accuracy of the predictions, we penalize the $2-$norm of $\alpha(t)$ in the cost function as well. This is because larger values of $\norm*{\alpha(t)}_2$ amplify the effect of the basis function approximation and noise in the data and, hence, smaller norms are preferred. Notice that when $\max\{\varepsilon^*,w^*\}\to0$, the optimal control problem \eqref{DDNMPC2} reduces to the nominal one in \eqref{DDNMPC} since (i) the regularization term of $\alpha(t)$ vanishes and (ii) the constraint \eqref{slack_const} results in $\norm*{\sigma(t)}_{\infty}\to0$. Finally, we assume that the input constraint set is compact and we do not consider output constraints i.e., $\mathcal{Y}=\mathbb{R}^m$, however, similar arguments can be made as in \cite{Berberich205} to ensure output constraint satisfaction. Algorithm~\ref{alg1} summarizes the proposed $d_{\textup{max}}-$step robust data-driven nonlinear predictive control scheme.
\begin{alg}\label{alg1}Robust data-driven nonlinear predictive control scheme.\\%
	1) At time $t$, use past $d_{\textup{max}}$ input-output measurements to solve the nonlinear optimization problem in \eqref{DDNMPC2}.\\
	2) Apply the optimal control input $\mathbf{\bar{u}}^*_{[0,d_{\textup{max}}-1]}(t)$ to \eqref{NLsys}.\\
	3) Set $t=t+d_{\textup{max}}$ and return to 1).			
\end{alg}%
To show practical exponential stability \cite[Def. 4.1]{Faulwasser18} and recursive feasibility of the proposed multi-step predictive control scheme, we require the following lemma which bounds the difference between the optimal predicted outputs $\bar{y}_{i,[0,L+d_i-1]}^*(t)$ and true outputs of the system $\breve{y}_{i,[t,t+L+d_i-1]}$ that result from applying $\mathbf{\bar{u}}^*(t)$ to System \eqref{NLsys}.
\begin{lemma}\label{lemma_output_difference}
	Let $\bar{\mathbf{u}}^*_{[0,L-1]}(t),\bar{y}_{i,[0,L+d_i-1]}^*(t),\alpha^*(t),\sigma^*(t)$ be solutions of \eqref{DDNMPC2} at time $t$, and let $\breve{\Xi}_{[t,t+L]}$ and $\breve{y}_{i,[t,t+L+d_i-1]}$ be the state and output of system \eqref{BINF} resulting from applying $\bar{\mathbf{u}}^*_{[0,L-1]}(t)$ to \eqref{NLsys} at time $t$. Then, for all $k\in\mathbb{Z}_{[0,L+d_i-1]}$, the following holds
	\begin{align}
		&\left|\breve{y}_{i,t+k} - \bar{y}_{i,k}^*(t)\right|\leq\mathcal{P}^{k + d_{\textup{max}} - d_i}(K_{\Xi})\Big(\varepsilon^*(1+\norm*{\alpha^*(t)}_1)\notag\\
		&\quad + (1 + K_w)w^*\norm*{\alpha^*(t)}_1 + (1+\norm*{\mathcal{G}}_{\infty})\norm*{\sigma^*(t)}_\infty\Big),
		\label{output_difference}
	\end{align}
	where $K_\Xi,K_w$ are defined in Remarks \ref{remark_Kw} and \ref{remark_KPsiKXi}, respectively, and $\mathcal{P}^k(K_\Xi)$ is the polynomial $\mathcal{P}^k(K_\Xi) = (K_\Xi)^k + (K_\Xi)^{k-1} + \dots + K_\Xi +1$.
\end{lemma}
\begin{pf}
		The outputs obtained when applying $\bar{\mathbf{u}}_{[0,L-1]}^*(t)$ to the system are given by
		\begin{equation}
			\breve{y}_{i,t+k} \stackrel{\eqref{y_for_proofs}}{=} \phi_i(\bar{\mathbf{u}}_{k-d_i}^*(t),\breve{\Xi}_{t+k-d_i}),
		\end{equation}
		for all $k\in\mathbb{Z}_{[0,L+d_i-1]}$, where $\bar{\mathbf{u}}_{[-d_i,-1]}^*(t)\stackrel{\eqref{pc2_ini}}{=}\mathbf{u}_{[t-d_i,t-1]}$. In contrast, the optimal predicted outputs are given by
		\begin{equation*}
			\begin{aligned}
				\bar{y}_{i,k}^*(t) &\stackrel{\eqref{pc2_willems},\eqref{Xi}}{=}H_1({\tilde{y}_{i,[k+d_{\textup{max}},k+N-L]}^{\textup{d}}})\alpha^*(t) - \sigma_{\Xi,k}^*(t)\\
				&= H_{1}({{y}_{i,[k+d_{\textup{max}},k+N-L]}^{\textup{d}}})\alpha^*(t)\\
				&\quad+ H_{1}({w_{i,[k+d_{\textup{max}},k+N-L]}^{\textup{d}}})\alpha^*(t)- \sigma_{\Xi,k}^*(t)
			\end{aligned}
		\end{equation*}
		\begin{equation*}
			\begin{aligned}
				\bar{y}_{i,k}^*(t)
				&\stackrel{\eqref{y_for_proofs}}{=}g_i^\top H_1\Big(\hat{\Psi}_{[k\hspace{-0.25mm}+d_{\textup{max}}\hspace{-0.25mm}-\hspace{-0.25mm}d_i,k\hspace{-0.25mm}+\hspace{-0.25mm}N\hspace{-0.25mm}-\hspace{-0.25mm}L\hspace{-0.25mm}-\hspace{-0.25mm}d_i]}(\mathbf{u}^{\textup{d}},\tilde{\Xi}^{\textup{d}})\\
				&\quad+\hat{E}_{[k\hspace{-0.25mm}+d_{\textup{max}}\hspace{-0.25mm}-\hspace{-0.25mm}d_i,k\hspace{-0.25mm}+\hspace{-0.25mm}N\hspace{-0.25mm}-\hspace{-0.25mm}L\hspace{-0.25mm}-\hspace{-0.25mm}d_i]}(\mathbf{u}^{\textup{d}},\tilde{\Xi}^{\textup{d}})\\
				&\quad+ \hat{D}_{[k\hspace{-0.25mm}+d_{\textup{max}}\hspace{-0.25mm}-\hspace{-0.25mm}d_i,k\hspace{-0.25mm}+\hspace{-0.25mm}N\hspace{-0.25mm}-\hspace{-0.25mm}L\hspace{-0.25mm}-\hspace{-0.25mm}d_i]}(\omega^{\textup{d}})\Big)\alpha^*\hspace{-0.25mm}(t)\\
				&\quad+ H_{\hspace{-0.5mm}1}({w_{i,[k\hspace{-0.25mm}+d_{\textup{max}}\hspace{-0.25mm},k\hspace{-0.25mm}+\hspace{-0.25mm}N\hspace{-0.25mm}-\hspace{-0.25mm}L]}^{\textup{d}}})\alpha^*\hspace{-0.25mm}(t)- \sigma_{\Xi,k}^*(t)\\
				&\stackrel{\eqref{pc2_willems},\eqref{imp_defs}}{=} g_i^\top \left(\Psi(\mathbf{\bar{u}}_{k-d_i}^*(t),\bar{\Xi}_{k-d_i}^*(t)) + \sigma_{\Psi,k-d_i}^*(t)\right)\\ 
				&\quad+ H_1(\hat{\varepsilon}_{i,[k+d_{\textup{max}}-d_i,k+N-L-d_i]}(\mathbf{u}^{\textup{d}},{{\Xi}^{\textup{d}}}))\alpha^*(t)\\
				&\quad + H_1(\hat{\delta}_{i,[k+d_{\textup{max}}-d_i,k+N-L-d_i]}(\omega^{\textup{d}}))\alpha^*(t) \\ 
				&\quad+ H_{\hspace{-0.5mm}1}({w_{i,[k\hspace{-0.25mm}+d_{\textup{max}}\hspace{-0.25mm},k\hspace{-0.25mm}+\hspace{-0.25mm}N\hspace{-0.25mm}-\hspace{-0.25mm}L]}^{\textup{d}}})\alpha^*\hspace{-0.25mm}(t)- \sigma_{\Xi,k}^*(t)\\
				&\stackrel{\eqref{basis}}{=} \phi_{i}(\mathbf{\bar{u}}_{k-d_i}^*(t),\bar{\Xi}_{k-d_i}^*(t)) - \hat{\varepsilon}_{i}(\mathbf{\bar{u}}_{k-d_i}^*(t),\bar{\Xi}_{k-d_i}^*(t))\\ 
				&\quad + g_i^\top\sigma_{\Psi,k-d_i}^*(t)- \sigma_{\Xi,k}^*(t)\\
				&\quad + H_1(\hat{\varepsilon}_{i,[k+d_{\textup{max}}-d_i,k+N-L-d_i]}(\mathbf{u}^{\textup{d}},{{\Xi}^{\textup{d}}}))\alpha^*(t) \\ 
				&\quad+ H_1(\hat{\delta}_{i,[k+d_{\textup{max}}-d_i,k+N-L-d_i]}(\omega^{\textup{d}}))\alpha^*(t)\\
				&\quad+ H_{\hspace{-0.5mm}1}({w_{i,[k\hspace{-0.25mm}+d_{\textup{max}}\hspace{-0.25mm},k\hspace{-0.25mm}+\hspace{-0.25mm}N\hspace{-0.25mm}-\hspace{-0.25mm}L]}^{\textup{d}}})\alpha^*\hspace{-0.25mm}(t).
			\end{aligned}
		\end{equation*}%
		The error between the two outputs for $k\in\mathbb{Z}_{[0,L+d_i-1]}$ is
			\begin{align}
				&\breve{y}_{i,t+k} - \bar{y}_{i,k}^*(t) = {\phi_i({\mathbf{\bar{u}}}^*_{k-d_i}(t),\breve{\Xi}_{t+k-d_i})}\label{phis_dont_cancel_out}\\
				&\quad - {\phi_{i}\left(\mathbf{\bar{u}}^*_{k-d_i}(t),\bar{\Xi}^*_{k-d_i}(t)\right)}+ \hat{\varepsilon}_{i}\left(\mathbf{\bar{u}}^*_{k-d_i}(t),\bar{\Xi}^*_{k-d_i}(t)\right)\notag \\
				&\quad -g_i^\top\sigma_{\Psi,k-d_i}^*(t) +\sigma_{\Xi,k}^*(t)\notag \\
				& \quad - H_1(\hat{\varepsilon}_{i,[k+d_{\textup{max}}-d_i,k+N-L-d_i]}(\mathbf{u}^{\textup{d}},{\tilde{\Xi}^{\textup{d}}}))\alpha^*(t)\notag\\
				&\quad- H_1(\hat{\delta}_{i,[k+d_{\textup{max}}-d_i,k+N-L-d_i]}(\omega^{\textup{d}}))\alpha^*(t)\notag\\
				& \quad- H_{\hspace{-0.5mm}1}({w_{i,[k\hspace{-0.25mm}+d_{\textup{max}}\hspace{-0.25mm},k\hspace{-0.25mm}+\hspace{-0.25mm}N\hspace{-0.25mm}-\hspace{-0.25mm}L]}^{\textup{d}}})\alpha^*\hspace{-0.25mm}(t).\notag
			\end{align}
		Using Assumption \ref{bounded_err_assmp}, and Lipschitz continuity of $\phi_i$, this can be bounded by
		\begin{align}
			&\left|\breve{y}_{i,t+k} - \bar{y}_{i,k}^*(t)\right|\leq K_{\Xi}\norm*{\breve{\Xi}_{t+k-d_i}-\bar{\Xi}_{k-d_i}^*(t)}_\infty\label{prel_lemma_part2_a}\\
			&\quad + \varepsilon^*(1+\norm*{\alpha^*(t)}_1) + (1 + K_w)w^*\norm*{\alpha^*(t)}_1\notag\\
			&\quad + (1+\norm*{\mathcal{G}}_{\infty})\norm*{\sigma^*(t)}_\infty.\notag
		\end{align}
		Using similar induction steps as in \cite[Theorem 3]{Alsalti2022a}, it can be shown that \eqref{prel_lemma_part2_a} implies
		\begin{align}
			&\left|\breve{y}_{i,t+k} - \bar{y}_{i,k}^*(t)\right|\leq\mathcal{P}^{k + d_{\textup{max}} - d_i}(K_{\Xi})\Big(\varepsilon^*(1+\norm*{\alpha^*(t)}_1)\notag\\
			&\quad + (1 + K_w)w^*\norm*{\alpha^*(t)}_1 + (1+\norm*{\mathcal{G}}_{\infty})\norm*{\sigma^*(t)}_\infty\Big),
			\label{prel_lemma_part2}
		\end{align}
		as in \eqref{output_difference} which completes the proof.\hfill$\blacksquare$
\end{pf}
\begin{remark}
Due to the polynomial $\mathcal{P}^{k}(K_{\Xi})$, the bound in \eqref{output_difference} increases with increasing $k$. Furthermore, if $K_{\Xi}<1$, then the bound converges as $k\to\infty$.
\end{remark}
Unlike Assumption \ref{PE_asmp}, we now require persistence of excitation of the sequence of basis functions evaluated using the noisy data, i.e., $\{\hat{\Psi}_k(\mathbf{u}^{\textup{d}},\tilde{\Xi}^{\textup{d}})\}_{k=0}^{N-1}$.
\begin{assumption}\label{noisy_PE_asmp}
	The input-output data are collected such that $\hat{\Psi}(\mathbf{u}^{\textup{d}},\tilde{\Xi}^{\textup{d}})$ is persistently exciting of order $L+d_{\textup{max}}+n$.
\end{assumption}
For sufficiently small $\varepsilon^*$ and $w^*$, Assumption \ref{noisy_PE_asmp} implies persistency of excitation of $\{\hat{\Psi}_k(\mathbf{u}^{\textup{d}},\tilde{\Xi}^{\textup{d}})+\hat{E}_k(\mathbf{u}^{\textup{d}},\tilde{\Xi}^{\textup{d}})+\hat{D}_k(\omega^{\textup{d}})\}_{k=0}^{N-1}$ of the same order. This is the case when, e.g., $\norm*{\hspace{-0.5mm}H_{L+d_{\textup{max}}}\hspace{-0.5mm}(\hat{E}(\mathbf{u}^{\textup{d}},\tilde{\Xi}^{\textup{d}})\hspace{-1mm}+\hspace{-1mm}\hat{D}(\omega^{\textup{d}}\hspace{-0.25mm})\hspace{-0.51mm})\hspace{-0.5mm}}_2\hspace{-2.5mm}<\hspace{-0.5mm}\sigma_{\textup{min}}(H_{L+d_{\textup{max}}}\hspace{-0.5mm}(\hat{\Psi}(\mathbf{u}^{\textup{d}},\tilde{\Xi}^{\textup{d}})))$. To show recursive feasibility, we use a candidate solution which is composed of appending the previously optimal solution by a deadbeat controller (see \cite{Alsalti2022b} for details). Therefore, the following assumption is imposed on the length of the prediction horizon.
\begin{assumption}\label{horizon_rob_asmp}
	The prediction horizon satisfies $L\geq d_{\textup{max}}$.
\end{assumption}
The following theorem states that the robust $d_{\textup{max}}$-step nonlinear predictive controller in \eqref{DDNMPC2} is recursively feasible and results in practical exponential stability of the closed-loop system.
\begin{thm}\label{rob_npc_thm}
	Let Assumptions~\ref{bounded_err_assmp}-\ref{horizon_rob_asmp} hold. Then, for any $V_{ROA}>0$, there exist $\underline{\lambda}_\alpha,\bar{\lambda}_\alpha,\underline{\lambda}_\sigma,\bar{\lambda}_\sigma>0$ such that for all $\lambda_\alpha, \lambda_\sigma>0$ satisfying
	\begin{equation*}
		\underline{\lambda}_\alpha \leq \lambda_\alpha \leq \bar{\lambda}_\alpha, \qquad \underline{\lambda}_\sigma \leq \lambda_\sigma \leq \bar{\lambda}_\sigma,
	\end{equation*}
	there exists constants $\bar{\varepsilon},\bar{w},\bar{c}_{\textup{pe}}>0$ as well as a continuous, strictly increasing function $\beta:[0,\bar{\varepsilon}]\times[0,\bar{w}]\to[0,V_{ROA}]$ with $\beta(0,0)=0$, such that for all $\varepsilon^*,w^*, c_{\textup{pe}}$ satisfying
	\begin{equation*}
		\varepsilon^*\leq \bar{\varepsilon},\qquad w^*\leq\bar{w},\qquad c_{\textup{pe}}\max\{\varepsilon^*,w^*\}\leq \bar{c}_{\textup{pe}},
	\end{equation*}
	there exist $c>0,\,P\succ0$ for the Lyapunov function $V_t=J^*_t+c\norm*{\Xi_t}_P^2$ such that the sub-level set $\mathbb{V}\coloneqq\{\Xi_t\in\mathbb{R}^n~|~V_t\leq V_{ROA}\}$ is invariant and $V_t$ converges exponentially to $V_t\leq \beta(\varepsilon^*, w^*)$ in closed-loop with the $d_{\textup{max}}-$step predictive control scheme for all initial conditions for which $V_0\leq V_{ROA}$.
\end{thm}
\begin{pf}
	The proof is provided in an online technical report and can be found in \cite{Alsalti2022b}. \hfill $\blacksquare$
\end{pf}
Theorem \ref{rob_npc_thm} shows that if the robust data-driven nonlinear predictive controller \eqref{DDNMPC2} is feasible at time $t$ with a Lyapunov function value less than or equal to $V_{ROA}>0$, then the Lyapunov function candidate converges to a neighborhood of the origin whose size $\beta(\varepsilon^*,w^*)$ depends on the basis functions approximation error bound $\varepsilon^*$ as well as the noise bound $w^*$. This, along with suitable upper and lower bounds on the Lyapunov function $V_t$, implies practical exponential stability of the closed-loop system as shown in detail in \cite{Alsalti2022b}. The set $\mathbb{V}$ of initial states for which $V_0\leq V_{ROA}$ serves as the guaranteed region of attraction of the closed-loop system. Furthermore, it is a ($d_{\textup{max}}$-step) invariant set and, hence, the closed-loop trajectories evolve in a compact subset of the input-state space in accordance with Assumption \ref{bounded_err_assmp}.\par
The data-based nonlinear predictive control schemes shown in \eqref{DDNMPC} and \eqref{DDNMPC2} involve solving a nonlinear optimization problem at each time instant. Solving this minimization problem to optimality can be computationally difficult to achieve. This is due to the nonlinear basis functions and the (potentially) large number of decision variables as the prediction horizon length and/or the number of basis functions $r$ increases. In total, the robust data-driven control scheme \eqref{DDNMPC2} has $N+(2m+r-1)(L+d_{\textup{max}})+n+1$ decision variables with $N\geq(r+1)(L+d_{\textup{max}}+n)-~1$, where the latter is required to ensure persistency of excitation of $\{\hat{\Psi}_k(\mathbf{u}^{\textup{d}},\tilde{\Xi}^{\textup{d}})\}_{k=0}^{N-1}$ of order $L+d_{\textup{max}}+n$. In this paper, we show simulation results on a model of a fully actuated double inverted pendulum. Applying this scheme experimentally is a topic for future research.
\subsection{Model-free bounds on $\norm*{\mathcal{G}}_{\infty}$}
Notice that the error bound \eqref{output_difference} requires knowledge of (an upper bound on) $\norm*{\mathcal{G}}_{\infty}$. The following lemma shows that, under certain conditions, one can obtain an upper bound of $\norm*{\mathcal{G}}_{\infty}$ in a model-free fashion. Furthermore, to implement the scheme in \eqref{DDNMPC2}, the value of $\norm*{\mathcal{G}^\dagger}_{\infty}$, or an upper bound on it\footnote{The guarantees of Theorem \ref{rob_npc_thm} remain true qualitatively if $\norm*{\mathcal{G}^\dagger}_\infty$ in \eqref{slack_const} is replaced by some upper bound, compare \cite{Alsalti2022b} for details. In particular, using an upper bound of $\norm*{\mathcal{G}^\dagger}_{\infty}$ only affects the values of some constants.} is required. In Remark \ref{remark_G} below, we explain how one can satisfy \eqref{slack_const} in practice using a sufficiently large, but finite, constant as an upper bound for $\norm*{\mathcal{G}^\dagger}_\infty$. 
\begin{lemma}\label{Lemma_G}
	Suppose Assumption~\ref{bounded_err_assmp} holds and let the basis functions $\psi_j$ be chosen such that the symmetric matrix of their inner products is invertible, i.e., det$(\Gamma)\neq0$ with $\Gamma_{j,\ell}=\left\langle \psi_j,\psi_\ell\right\rangle$. Then, there exists $v^*>0$ such that
	\begin{equation}
		\norm*{\mathcal{G}}_{\infty} \leq v^*\norm{\Gamma^{-1}}_1\sum\limits_{j=1}^r\int_{\Omega}|\psi_j(s_1,s_2)|ds_1ds_2.\label{G_bnd_a}
	\end{equation}
\end{lemma}
\begin{pf}
	By the definition of $\mathcal{G}$ in \eqref{def_of_G_mat}, each row $g_i^\top$ satisfies
	\begin{align}
		g_i{=}\arg\min\limits_{g_i}\hspace{0mm}\left< \phi_{i},\phi_{i}\right>\hspace{0mm}- 2\left<\phi_{i},g_i^\top{\Psi}\right>+\left< g_i^\top{\Psi},g_i^\top{\Psi}\right>.\label{min_for_g}
	\end{align}
	Since $\Gamma$ is assumed to be invertible, \eqref{min_for_g} represents a least squares problem whose unique solution can be obtained from $\Gamma {g_i} = \zeta$, where
	\begin{equation*}
		\begin{aligned}
			\Gamma_{j,\ell} = \left<{\psi}_j,{\psi}_\ell\right>,\qquad \zeta_{j} = \left<\phi_i,{\psi}_j\right>.
		\end{aligned}
	\end{equation*}
	Therefore, the solution to \eqref{min_for_g} is $g_i = \Gamma^{-1}\zeta$ and, hence,
	\begin{equation}
		\norm*{g_i^\top}_\infty = \norm{g_i}_1 \leq \norm{\Gamma^{-1}}_1 \norm{\zeta}_1.\label{bound_for_gi_proof_step1}
	\end{equation}
	Moreover, using the definition of $\zeta$, we write
	\begin{equation*}
		\begin{aligned}
			\norm{\zeta}_1 &= \norm*{\begin{bmatrix}
					\int_{\Omega}\phi_i(s_1,s_2){\psi}_1(s_1,s_2)ds_1ds_2\\
					\vdots\\
					\int_{\Omega}\phi_i(s_1,s_2){\psi}_r(s_1,s_2)ds_1ds_2\\
			\end{bmatrix}}_1\\
			&\leq\sum\limits_{j=1}^r\int_{\Omega}\left|\phi_i(s_1,s_2){\psi}_j(s_1,s_2)\right|ds_1ds_2.
		\end{aligned}
	\end{equation*}
	By the extreme value theorem, the function $\phi_i$ (which is locally Lipschitz continuous) attains a maximum on the compact set $\Omega$. Let the maximum be $v_i^*$ and write
	\begin{equation*}
		\norm{\zeta}_1 \leq v_i^*\sum\limits_{j=1}^r\int_{\Omega}\left|{\psi}_j(s_1,s_2)\right|ds_1ds_2.
	\end{equation*}
	Plugging back in \eqref{bound_for_gi_proof_step1} results in
	\begin{equation*}
		\norm*{g_i}_1 \hspace{-0.75mm} =\hspace{-0.75mm} \norm*{g_i^\top}_\infty \hspace{-0.75mm}\leq\hspace{-0.75mm} v_i^* \norm*{\Gamma^{-1}}_1\hspace{-1mm} \sum\limits_{j=1}^r\hspace{-0.5mm}\int_{\Omega}\hspace{-1mm}\left|{\psi}_j(s_1,s_2)\right|ds_1ds_2.
	\end{equation*}
	 Finally, since $\norm*{\mathcal{G}}_\infty = \max_i\norm*{g_i^\top}_{\infty}$ we obtain \eqref{G_bnd_a} with $v^*=\max_i v_i^*$, which completes the proof. \hfill $\blacksquare$
\end{pf}
\begin{remark}
	Invertibility of the matrix $\Gamma$ in Lemma \ref{Lemma_G} is guaranteed if the basis functions are orthogonal in $\Omega$.
\end{remark}
\begin{remark}\label{remark_G}
	Since $\mathcal{G}\mathcal{G}^\dagger=I$, it holds that $\norm*{\mathcal{G}^\dagger}_{\infty}\leq\frac{\sqrt{r}}{\sigma_{\textup{min}}(\mathcal{G})}$. Since $\mathcal{G}$ has full row rank by assumption, $\frac{\sqrt{r}}{\sigma_{\textup{min}}(\mathcal{G})}$ is a finite number (since $\sigma_{\textup{min}}(\mathcal{G})\neq0$). For implementing the scheme in \eqref{DDNMPC2}, one can use a large constant in order to satisfy \eqref{slack_const}. Following similar arguments as in \cite[Remark 3]{Berberich203} and \cite{Bongard21}, one can show that the constraint in \eqref{slack_const} can be dropped by suitably adapting the regularization parameter in the cost function.
\end{remark}
	\section{Example}\label{examples}
In this section, we implement the robust nonlinear predictive control scheme in Algorithm \ref{alg1} to stabilize a {nonzero} set point of a discretized model of a fully-actuated double inverted pendulum, despite using an inexact basis function approximation of the unknown nonlinearities as in \eqref{basis}. A discrete-time model of this system can be obtained using Euler's discretization of the continuous-time dynamics, which results in
\begin{align}
		&\mathbf{x}_{k+1} = \mathbf{x}_k + T_s\left(\mathcal{A}\mathbf{x}_k + \mathcal{B}Z_k\right),\quad \mathbf{y}_k = \mathcal{C} \mathbf{x}_k,\label{linearizing_controller_ct}\\
		&\mathbf{x}_k\coloneqq \begin{bmatrix}\theta_{1,k} & \vartheta_{1,k} & \theta_{2,k} & \vartheta_{2,k}\end{bmatrix}^\top, \quad \mathbf{y}_k=\begin{bmatrix}\theta_{1,k} & \theta_{2,k}\end{bmatrix}^\top,\notag\\
		&Z_k \hspace{-1mm}\coloneqq\hspace{-1mm}\begin{bmatrix}z_{1,k}\\ z_{2,k}\end{bmatrix} \hspace{-1mm}=\hspace{-1mm} M(\theta_k)^{-1}\hspace{-1mm}\left(\tau_k -\hspace{-0.5mm} C(\theta_k,\vartheta_k)\vartheta_k \hspace{-0.75mm}-\hspace{-0.75mm} G(\theta_k)\hspace{-0.25mm}\right)\hspace{-0.5mm}.
		\notag
\end{align}
{In \eqref{linearizing_controller_ct}, $\theta_k,\vartheta_k,\tau_k$ are the vectors of angular positions, angular velocities and joint torques respectively, and $T_s$ is the sampling time. The terms $M(\theta),\,C(\theta,\vartheta)$ and $G(\theta)$ represent the inertia, dissipative and gravitational terms, respectively, and depend on the masses and lengths of the two links\footnote{The following (unknown) model parameters were used: $m_1=m_2=1$kg and $l_1=l_2=0.5$m for the masses and lengths of the two links, respectively.} $m_1,m_2,l_1,l_2$ \cite{spong20}. The outputs in \eqref{linearizing_controller_ct} have relative degrees $d_1=d_2=2$ and $\sum_id_i=4=n$. Thus, as in Section \ref{sec_MBFLsys}, there exists a coordinate transformation $\Xi_k=T(\mathbf{x}_k)$ such that the transformed system is full-state feedback linearizable. This transformation takes the form
	\begin{equation}
		\Xi_k = \begin{bmatrix}
			x_{1,k}, & x_{1,k}+T_sx_{2,k}, & x_{3,k}, & x_{3,k} + T_sx_{4,k}
		\end{bmatrix}^\top.\label{coordinate_transformation}
	\end{equation}
	Re-writing \eqref{linearizing_controller_ct} in the transformed coordinates results in
	\begin{gather}
		\begin{matrix}
			\Xi_{k+1} = \mathcal{A}\Xi_k + \mathcal{B}\mathbf{v}_k,&\qquad 
			\mathbf{y}_k=\mathcal{C}\Xi_k,
		\end{matrix}\label{dt_bruno}\\
		\mathbf{v}_k \coloneqq \begin{bmatrix}
			v_{1,k} \\ v_{2,k}
		\end{bmatrix} = \begin{bmatrix}
			2\xi_{2,k} - \xi_{1,k} + T_s^2z_{1,k} \\
			2\xi_{4,k} - \xi_{3,k} + T_s^2z_{2,k}
		\end{bmatrix}.\label{linearizing_controller_dt}
\end{gather}}%
Notice that for the system in \eqref{dt_bruno}, there is no choice of {parameter-independent} basis functions that span $\mathbf{v}_k$. This is because $Z_k$ in \eqref{linearizing_controller_dt} depends on $M(\theta)^{-1}$, which is a matrix consisting of model parameters. Therefore, in order to exactly reconstruct the function $\mathbf{v}_k$ from any combination of basis functions, knowledge of model parameters is required. For the case where model parameters are not exactly known and the output data is noisy, we apply the results of Theorem \ref{rob_npc_thm} to stabilize the system using the robust data-driven nonlinear predictive control scheme \eqref{DDNMPC2}. In particular, we will use the following choice of basis functions to approximate $\mathbf{v}_k$ in \eqref{linearizing_controller_dt}
\begin{align}
	&\Psi(\tau_k,\Xi_k) =\label{choice_of_Psi}\\ 
	&\begin{bmatrix}
		\tau_k\\
		\hspace{-1mm}{\tilde{M}\hspace{-0.5mm}\left(\hspace{-0.5mm}\begin{bmatrix}
				\xi_{1,k}\\ \xi_{3,k}
			\end{bmatrix}\hspace{-0.5mm}\right)}^{\hspace{-1mm}-1}\hspace{-2mm}\left(\hspace{-0.5mm} \tau_k \hspace{-0.5mm}- \hspace{-0.5mm}\tilde{C}(\Xi_k)\hspace{-0.5mm}\begin{bmatrix}
			(\xi_{2,k} - \xi_{1,k}) / T_s\\ (\xi_{4,k} - \xi_{3,k}) / T_s
		\end{bmatrix}\hspace{-0.5mm} - \hspace{-0.5mm}\tilde{G}\hspace{-0.5mm}\left(\hspace{-0.5mm}\begin{bmatrix}
			\xi_{1,k}\\ \xi_{3,k}
		\end{bmatrix}\hspace{-0.5mm}\right)\hspace{-0.5mm}\right)\hspace{-1mm}
	\end{bmatrix}\notag
\end{align}
where $\tilde{M},\,\tilde{C},\,\tilde{G}$ contain user-provided estimates for parameter values (in this example, they were obtained by randomly perturbing the unknown real values by up to 10\%). Notice that we exploit nominal model {structure} in the choice of basis functions. This is justified by the fact that the model structure in \eqref{linearizing_controller_ct} is ubiquitous in robotics, unlike the model parameters which can be very difficult to obtain and are essential for model-based robot control techniques like, e.g., impedance control or computed torque method \cite{spong20}.\par
\begin{figure*}[t]
	\centering\includegraphics[width=0.975\textwidth]{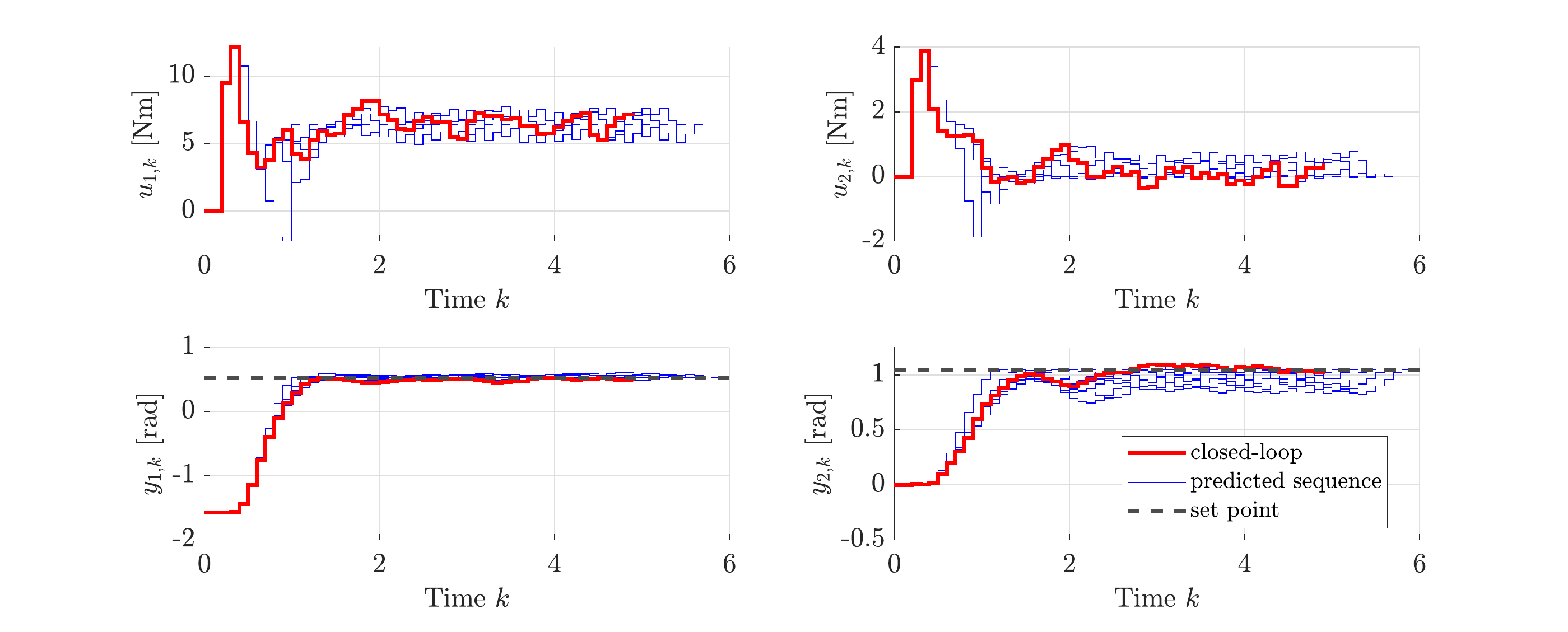}
	\caption{Swing up and practical stabilization of a non-zero set point of the double-inverted pendulum using the proposed data-driven nonlinear predictive controller in Theorem \ref{rob_npc_thm}.}
	\label{fig2_ex2}
\end{figure*}
We collect persistently exciting input-output data of length $N=200$ (or 20 seconds with a sampling time $T_s=0.1$) by operating the double inverted pendulum using a pre-stabilizing controller within the following compact subset of the input-state space
\begin{equation*}
	\Omega = \left \lbrace (\tau_k,{\Xi}_k)\in\mathbb{R}^2\times\mathbb{R}^4 ~\left|~ \begin{bmatrix}
		\tau_{\textup{lb}} \\ \Xi_{\textup{lb}}
	\end{bmatrix} \hspace{-0.5mm}\leq \hspace{-0.5mm}\begin{bmatrix}
		\tau_k\\ {\Xi}_k
	\end{bmatrix} \hspace{-0.5mm}\leq \hspace{-0.5mm}\begin{bmatrix}
		\tau_{\textup{ub}} \\ \Xi_{\textup{ub}}
	\end{bmatrix} \right. \right \rbrace,
\end{equation*}
where the inequalities are defined element-wise and $\tau_{\textup{ub}} = -\tau_{\textup{lb}} = \begin{bmatrix}
	20 & 20
\end{bmatrix}^\top$ Nm, $\Xi_{\textup{ub}}= -\Xi_{\textup{lb}} =\begin{bmatrix}
\pi/2 & \pi/2 & \pi/2 & \pi/2
\end{bmatrix}^\top$. In $\Omega$, the basis functions approximation error is upper bounded by\footnote{This was obtained by gridding $\Omega$ and numerically solving for $\mathcal{G}$.} $\varepsilon^*= 1.2893$ as in Assumption \ref{bounded_err_assmp}, whereas the measurement noise was upper bounded by $w^*=0.01$. The control objective is to stabilize the output set point $y_1^s=\pi/6\textup{ rad},\,y_2^s=\pi/3\textup{ rad}$ with the corresponding known input set point $u_1^s=6.3718 \textup{ Nm},\,u_2^s=0 \textup{ Nm}$. We use quadratic stage costs as in \eqref{stage_cost_fcn} and set the weighting matrices to $Q=R=I_2$. Furthermore, the prediction horizon was set to $L=10$. Finally, the regularization parameters in \eqref{opt_val_fcn_def} were set to $\lambda_\alpha=10^4,\, \lambda_\sigma=10^8$, while the input constraint set \eqref{rob_input_const_set} is $\mathcal{U}=\{\tau_k\in\mathbb{R}^2~\left|~ \tau_{\textup{lb}}\leq\tau_k\leq\tau_{\textup{ub}}\right.\}$.
\begin{remark}
	In order to implement the robust data-driven scheme in \eqref{DDNMPC2}, knowledge of $K_\Psi,K_w$ and $\norm*{\mathcal{G}^\dagger}_{\infty}$ are required. However, one can relax this constraint by enforcing $\norm*{\sigma_k(t)}_{\infty}\leq c\max\{\varepsilon^*,w^*\}$ for a sufficiently large constant $c>0$ and still retain the same theoretical guarantees as in Theorem~\ref{rob_npc_thm} (cf. \cite[Remark 3]{Berberich203} and \cite{Bongard21}). In this example, this relaxation was implemented.
\end{remark}
Figure \ref{fig2_ex2} shows the results of applying the nonlinear predictive control scheme when starting from the downward position of the pendulum. As seen in Figure \ref{fig2_ex2}, the robust nonlinear predictive control scheme manages to practically stabilize the given set point despite inexact basis function decomposition and output noise. Depending on the collected data, it was observed that longer sequences of previously collected data may yield better performance, but require more computation time at each time step $t$ since the size of the Hankel matrices in \eqref{pc2_willems} increases. Finally, it was observed that the steady-state error decreases when the noise level decreases and when the estimates $\tilde{M},\tilde{C},\tilde{G}$ get better (i.e., correspond to smaller $\varepsilon^*$), which agrees with the results of Theorem \ref{rob_npc_thm}.
	\section{Conclusions}\label{conclusion_sec}
In this paper, we presented a data-driven robust predictive control scheme for DT-MIMO feedback linearizable nonlinear systems. We formally showed that this scheme is recursively feasible and leads to practical exponential stability of the closed-loop system under bounded inexact basis function approximation error and output measurement noise. Finally, the results were analyzed on a discretized model of a fully-actuated double inverted pendulum. The results show that the scheme has good inherent robustness properties but can potentially be computationally expensive to implement, since the complexity of the nonlinear optimal control problem increases as more (complex) basis functions are used. Venues for future research are to extend this scheme to larger classes of nonlinear systems and to conduct a more comprehensive case study of an experimental implementation of the proposed scheme.
	
	{\bibliography{ifacconf}}
	
\end{document}